\documentclass[11pt]{article}
\usepackage{float} 
\usepackage[symbol]{footmisc} 
\usepackage[superscript,sort]{cite}
\usepackage{array}
\usepackage{bm}
\usepackage{setspace}
\usepackage{lineno}
\usepackage[normalem]{ulem} 
\usepackage[colorlinks=true,linkcolor=blue,citecolor=black,urlcolor=black,bookmarks=true,breaklinks=true]{hyperref}
\usepackage[usenames]{color}
\usepackage{amsmath,amssymb}
\usepackage{amsthm,amscd,amsxtra,amsfonts,amsmath,amssymb,multirow}
\usepackage{wrapfig}
\usepackage[footnotesize]{caption}
\usepackage{subcaption}
\usepackage{threeparttable} 
\usepackage{helvet}

\usepackage{booktabs}
\usepackage[table]{xcolor}
\usepackage{xcolor,colortbl}
\usepackage{placeins}
\usepackage{graphicx}
\usepackage{comment}
\usepackage[T1]{fontenc}
\newcommand{\beq}[1]{\begin{equation} \label{#1}}
\newcommand{\eeq}{\end{equation}}
\newcommand{\barray}{\begin{array}{ll}}
\newcommand{\earray}{\end{array}}

\title{Topological and geometric analysis of cell states in single-cell transcriptomic data}
\author{Tram Huynh$^1$
and
Zixuan Cang$^{1}$ \footnote{Corresponding author. Email: zcang@ncsu.edu}
\\
$^1$ Department of Mathematics and Center for Research in Scientific Computing, \\
North Carolina State University, NC 27695, USA}
\date{}

\begin{document}

\maketitle
\begin{abstract}
Single-cell RNA sequencing (scRNA-seq) enables dissecting cellular heterogeneity in tissues, resulting in numerous biological discoveries. Various computational methods have been devised to delineate cell types by clustering scRNA-seq data where the clusters are often annotated using prior knowledge of marker genes. In addition to identifying pure cell types, several methods have been developed to identify cells undergoing state transitions which often rely on prior clustering results. Present computational approaches predominantly investigate the local and first-order structures of scRNA-seq data using graph representations, while scRNA-seq data frequently displays complex high-dimensional structures. Here, we present a tool, scGeom for exploiting the multiscale and multidimensional structures in scRNA-seq data by inspecting the geometry via graph curvature and topology via persistent homology of both cell networks and gene networks. We demonstrate the utility of these structural features for reflecting biological properties and functions in several applications where we show that curvatures and topological signatures of cell and gene networks can help indicate transition cells and developmental potency of cells. We additionally illustrate that the structural characteristics can improve the classification of cell types.
\end{abstract}

\section{Introduction}
Single-cell RNA sequencing (scRNA-seq) provides a high-throughput measurement of gene expression profiles of individual cells which enables dissecting cellular heterogeneity in an unprecedented resolution\cite{svensson2018exponential}. Many computational methods have been developed for scRNA-seq data and have revealed numerous novel cell types, and differentiating trajectories\cite{luecken2019current}. Clustering and trajectory inference are two main analysis tasks. They are often performed on a reduced dimensional space in which there is a metric to describe similarity among cells. In clustering, each cell cluster potentially represents a cell type and is often annotated by confirming marker genes with prior knowledge. In trajectory inference, a graph is often constructed by connecting cells with similar gene expression profiles upon which minimal spanning trees or graph coarsening can be performed to summarize the trajectory structures. Identifying transition cells between states is crucial for inferring the local transitions between stable cell states. Compared to cells within a single cell type, transition cells between cell types are often not as effectively captured in scRNA-seq data due to the instability of transition states. Moreover, the biological properties of cell types such as developmental potency are mostly annotated using prior knowledge\cite{ashburner2000gene}. Predicting the developmental potency can annotate the global temporal directionality in a dataset. Computational methods for exploring transition states and unsupervised analysis of developmental potency remain understudied.

Recently, several methods have been developed to study the transition states between cell types. QuanTC\cite{sha2020inference} and scRCMF\cite{zheng2019scrcmf} perform non-negative matrix factorizations on the cell-by-cell similarity matrix and cell-by-gene expression matrix, respectively with each factor representing a cell type. The entropy of the assignment scores of each cell to the cell types is used as an indicator of transition cells. Soft clustering algorithms can also derive soft assignment scores to determine the pure and transition cells such as SOUP\cite{zhu2019semisoft}, DBCTI\cite{lan2022density} and scTite\cite{gan2022entropy}, using predefined criteria or entropy. MuTrans\cite{zhou2021dissecting} models scRNA-seq data as a dynamical system based on a cell-fate dynamical manifold determined from clustering and can identify transition cells also using the entropy of assignment scores of cells to sinks. Cabybara\cite{kong2022capybara} utilizes the vast reference databases of annotated bulk and single-cell transcriptomic data to assign cell types to the single cells and identifies cells predicted to have hybrid cell types as transition cells. These methods depend on clustering of data which often requires a predefined number of clusters or classification of data using reference training data. Here, we aim to explore the rich structures underlying single-cell data to infer transition cells without depending on clustering or classification results.

Analyzing pluripotency or developmental potency of cell types is valuable for refining structures and assigning global directions to the pseudo-temporal trajectories inferred from scRNA-seq data. With the accumulation of large-scale networks such as gene regulatory networks and protein-protein interaction networks, and computational methods for inferring large-scale gene networks from scRNA-seq data such as the correlation-based ones\cite{dai2019cell,wang2021constructing}, there exists an opportunity to infer the pluripotency by examining the structures of gene networks. For example, entropy\cite{banerji2013cellular,teschendorff2017single} and curvature\cite{murgas2022hypergraph} on gene networks have been used to reflect the pluripotency of cells. These methods use global summaries of the local properties of the gene networks. Here, we aim to further use topological methods for multiscale exploration of both local and global structures of the gene networks.

The high-dimensional scRNA-seq data assembles a complex heterogeneous manifold, while the emerging field of topological and geometric data analysis (TGDA)\cite{wasserman2018topological} particularly aims to systematically extract structural information from such complex structures. Mapper\cite{singh2007topological} is one of the major tools in TGDA that derives a structural abstraction of often high-dimensional data and has been applied to scRNA-seq data for extracting a simplified manifold underlying the data\cite{rizvi2017single}. Another major tool, persistent homology\cite{zomorodian2004computing,edelsbrunner2002topological,edelsbrunner2008persistent}, systematically examines topological features of different dimensions and at various geometric scales. Persistent homology has found its applications in various biological fields such as analyzing neural activity data\cite{sizemore2019importance} and structure-based biomolecular property predictions\cite{cang2017topologynet,meng2020weighted}. Persistent homology is generally applicable for different types of data including point clouds, volumetric data\cite{kaczynski2004computational}, and networks\cite{aktas2019persistence}. Its application in scRNA-seq data, however, remains unexplored.

Here, we aim to explore the usage of TGDA tools, specifically graph curvature and persistent homology, for establishing structure-function relationships in scRNA-seq to predict cell properties from the underlying structures of the data. We focus on two types of structures, a network of cells with cells connected based on their gene expression similarities and gene networks associated with each cell. Based on the cell network, we use Ollivier-Ricci curvature, a discretization of Ricci curvature on graphs, local persistent homology and relative persistent homology to identify transition cells independent from clustering or classification of cell types. For gene networks, we use vertex-based clique complex and edge-weighted Vietoris-Rips complex-based persistent homology to characterize node-weighted knowledge-based gene networks and edge-weighted cell-specific gene networks, respectively. The topological summaries are then related to the pluripotency or developmental potency of the cells. In a more general case, we also explore the usage of topological summaries as additional features in the task of cell type classification. These utilities are demonstrated on several real datasets with ground truth from scRNA-seq data on real time points or expert annotations. 

\section{Results}
\subsection{Method Overview}
\FloatBarrier

\begin{figure}[h]
\includegraphics[width=0.65\textwidth]{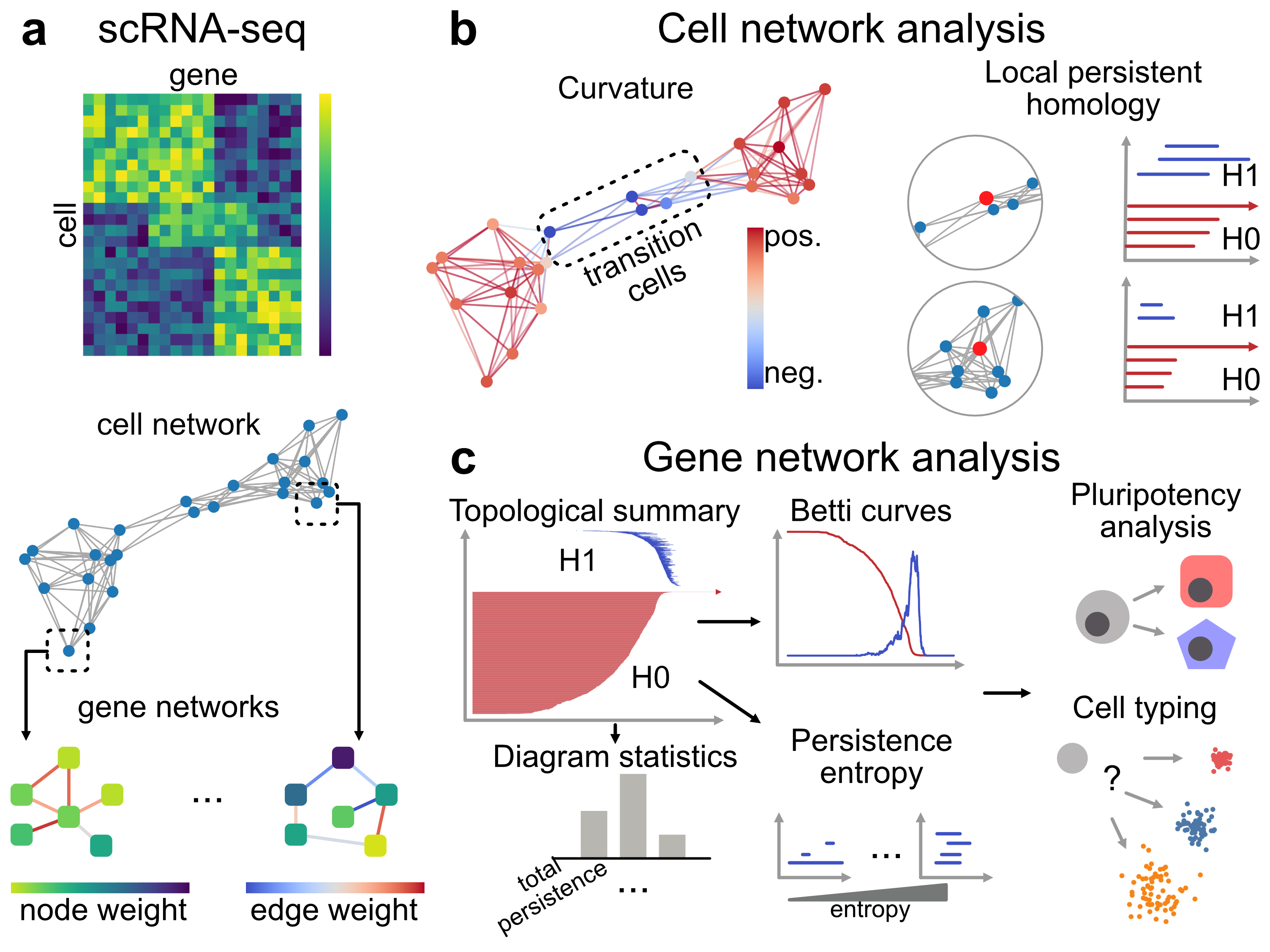}
\centering
\caption{\textbf{Overview of scGeom.} \textbf{a} The structure of a scRNA-seq data is often represented as cell networks where the cell-specific gene networks can be inferred for the cells. \textbf{b} The local structure of each cell is described by curvatures and local topology which are correlated to cell states. \textbf{c} The structures of cell-specific gene networks are characterized by various topological descriptors that are used to link to cell properties such as pluripotency and cell types.}
\label{fig:overview}
\end{figure}

To explore the structure-function relationship underlying single-cell data, we develop scGeom, a tool that characterizes the geometric and topological properties of cell networks and gene networks and relate them to the biological properties of cells (Fig. \ref{fig:overview}). The single-cell data is first preprocessed following the common pipelines of normalization, highly variable gene selection and PCA dimension reduction. A cell network denoted by $G_\mathrm{c}=(V_\mathrm{c},E_\mathrm{c})$ is then constructed by building a $k$-nearest neighbor graph with respect to Euclidean distance of the PCA embeddings. On $G_\mathrm{c}$, a graph curvature is computed for each edge using a discretization of Ricci curvature on graphs, the Ollivier-Ricci curvature (ORC)\cite{ollivier2009ricci} which measures the divergence of local geometry from the Euclidean space. Specifically, for the edge $e_{ij}\in E_\mathrm{c}$, ORC examines the difference between the edge length $e_{ij}$ and the distance between neighborhoods of $v_i$ and $v_j$ by optimal transport with shortest path distance as the ground cost. The curvatures on nodes are then defined by summing over the corresponding edges. A cell within a community or between communities is likely to have a positive or negative graph curvature respectively, analogous to scalar curvature in Riemannian geometry. 

The topological structures are characterized by persistent homology\cite{edelsbrunner2002topological,zomorodian2004computing}. The structure of the network is represented by a growing sequence of simplicial complexes that generalize graphs to higher dimensions. This sequence is called a filtration which captures the structural features at various scales compared to using one fixed graph or simplicial complex. Along the filtration, persistent homology tracks the appearances and disappearances of $k$-dimensional holes and their persistence through the filtration. For example, the $0$, $1$, and $2-$dimensional holes correspond to connected components, loops, and voids. Given a graph or point cloud, persistent homology outputs collections of persistence intervals, also called persistence diagrams, $\mathrm{Dgm}_k=[b_i,d_i)_{i=1}^{n^{(k)}}$ representing the filtration values corresponding to the appearance ($b_i$) and the disappearance ($d_i$) of the $k$th homology groups associated to the $k$-dimensional holes. Details of graph curvature and persistent homology are discussed in Sections \ref{sec:graphcurvature} and \ref{sec:persistenthomology}. For each cell in the cell network, persistent homology is computed for its local neighborhood. The graph curvature $\kappa$ and featurizations of the persistence diagram such as the total persistence $\sum_i(d_i-b_i)$ are used to distinguish cells in the transition and stable states (Fig. \ref{fig:overview}b). 

A cell-specific gene network\cite{dai2019cell}, $G_\mathrm{g}$ is constructed to reveal higher-order properties of cells in addition to the first-order gene expression levels. Persistence diagrams are computed for $G_\mathrm{g}^i$ of cell $i$ using filtrations such as edge-weighted Vietoris-Rips complex, resulting in persistence diagrams $\mathrm{Dgm}_k(G_\mathrm{g}^i)$. We then turn the persistence diagrams into features including persistence images that fit kernels to the $\mathrm{Dgm}_k$ regarded as point clouds, Betti curves that count the number of homology groups at every filtration value, and various statistics of $\mathrm{Dgm}_k$ such as total persistence and longest persistence. These features are then used to analyze the pluripotency or developmental potency of cells and fed to machine learning methods together with gene expression features for predicting cell types (Fig. \ref{fig:overview}c). Details of the methods and preprocessing of data can be found in Section \ref{sec:method}.

\subsection{Identifying transition cells with curvature and local topology}

\begin{figure}[h]
\includegraphics[width=0.7\textwidth]{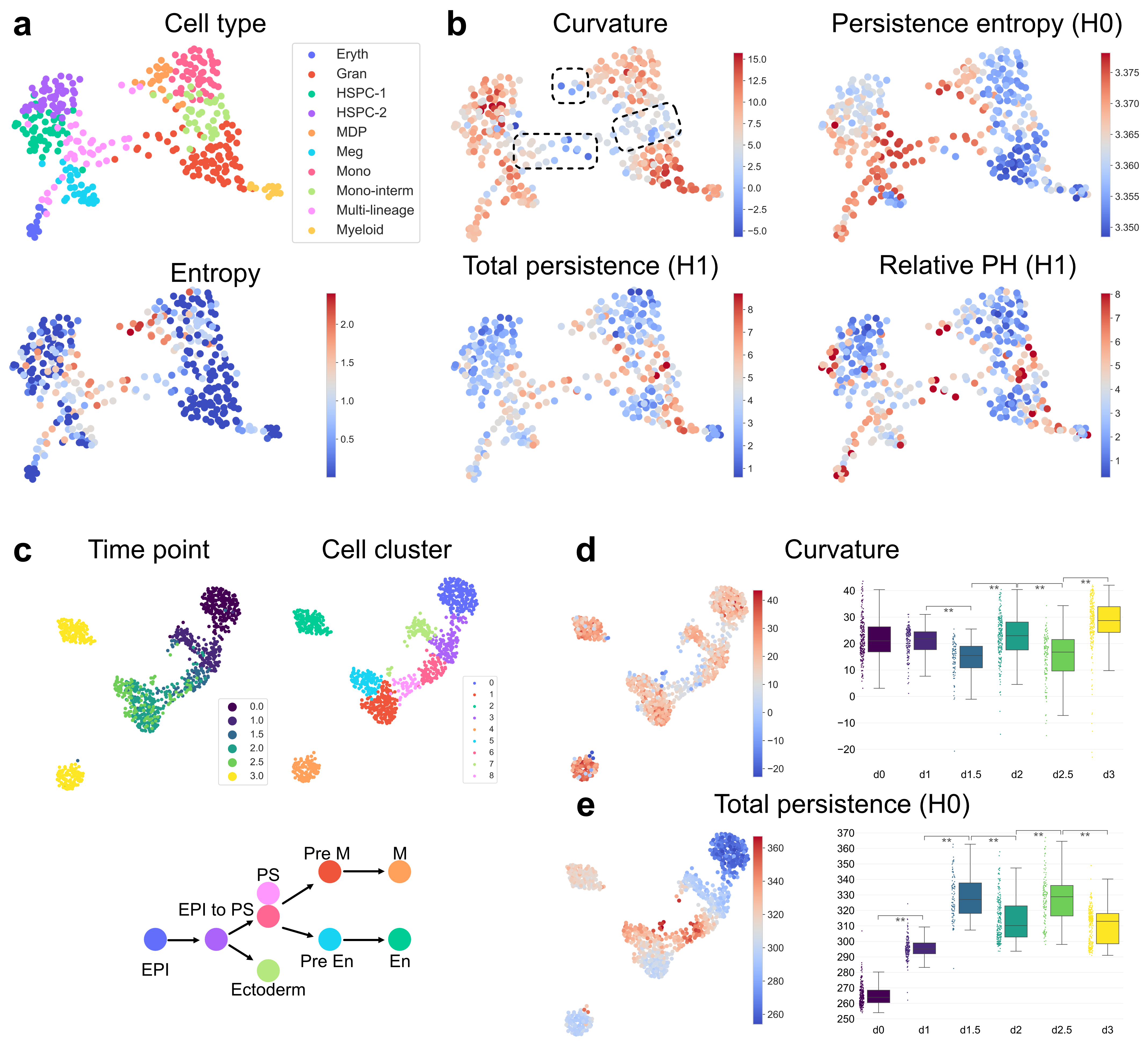}
\centering
\caption{\textbf{Analysis of transition states.} \textbf{a} A single-cell dataset of myelopoiesis. Applying MuTrans results in the entropy measuring the uncertainty of cluster assignment of a cell which indicates transition cells. \textbf{b} The structural features in the myelopoiesis dataset computed by scGeom on the cell network for each cell including curvature, local persistent homology, and relative persistent homology. The local persistent homology summarized as persistence entropy and total persistence, and the relative persistent homology described by the Wasserstein distance between relative persistence diagram and regular persistence diagram are shown. \textbf{c} A single-cell dataset of induced pluripotent stem cells (iPSC) taken at several temporal points from day 0 to day 3 where two transition events happen at day 1.5 and day 2.5. \textbf{d,e} The curvature on the cell network and the total persistence of the local persistent homology output. ** indicates $p$-value less than 1e-10 by Wilcoxon test.}
\label{fig:transition}
\end{figure}

We first analyzed a scRNA-seq data of myelopoiesis which captures several transitional intermediate states during differentiation of blood cells \cite{olsson2016single}. In this dataset, two relatively unstable states, a multi-lineage state and a monocyte intermediate state were identified by the original study using the ICGS approach\cite{olsson2016single} and another analysis of the dataset using a physics-based modeling tool MuTrans\cite{zhou2021dissecting} (Fig. \ref{fig:transition}a). Further, the MuTrans analysis constructed a differentiation landscape and identified transition cells between states depicted by the entropy of probability score assigned to each state (Fig. \ref{fig:transition}a). Here, based on $k$-nearest neighbors graphs of cells using the PCA embedding, we computed the Ollivier-Ricci curvature for each cell. We also computed topological features of each cell including local persistent homology on a neighborhood graph centered at the cell and the relative persistent homology of the global structure relative to a small neighborhood of the cell. The local persistent homology captures the multiscale and multidimensional structural characteristics of the local structure centered at each cell and the resulting persistence diagrams are turned into features by computing the total persistence and the persistence entropy\cite{atienza2020stability}. The relative persistent homology examines the significance of a cell in defining the global structure of the dataset. The resulting persistence diagrams are described by computing the Wasserstein distance between the relative persistence diagram and the regular persistence diagram of the whole dataset. These geometric and topological features were able to highlight both the relatively unstable cell states and the transition cells between states (Fig. \ref{fig:transition}b).

We then analyzed a single-cell dataset of induced pluripotent stem cells (iPSCs) taken at several real time points\cite{bargaje2017cell} (Fig. \ref{fig:transition}c). This dataset depicts two major transition events, epiblast (EPI) to primitive-streak (PS) cells and PS to mesenchymal (M) and endodermal (En) cells. Utilizing biological knowledge, the original study\cite{bargaje2017cell} projected these two transition events to happen around day 1.5 and day 2.5, respectively. Here, we perform an unbiased analysis without using any prior knowledge. We computed the graph curvature and local persistent homology on the PCA embedding and found a significant decrease in curvature and an increase in total persistence at day 1.5 and day 2.5. A smaller curvature indicates bridges between stable states. For topology, more significant topological features, for example, higher total persistence, reflect divergence from trivial structures and thus reveal the transition processes. 

\subsection{Topological signature reflects developmental potential}
In addition to examining the structures of cell networks, we further explore the relation between cell states and the structures of gene networks. A prior knowledge-based gene network\cite{banerji2013cellular} was assigned to each cell where the node weights were determined by the gene expression levels in the corresponding cell. For each cell, persistent homology was computed on this node-weighted gene network using vertex-based clique complex filtration where the edge filtration value is determined as the smaller weights of its two nodes. 
\begin{figure}[h]
\includegraphics[width=0.85\textwidth]{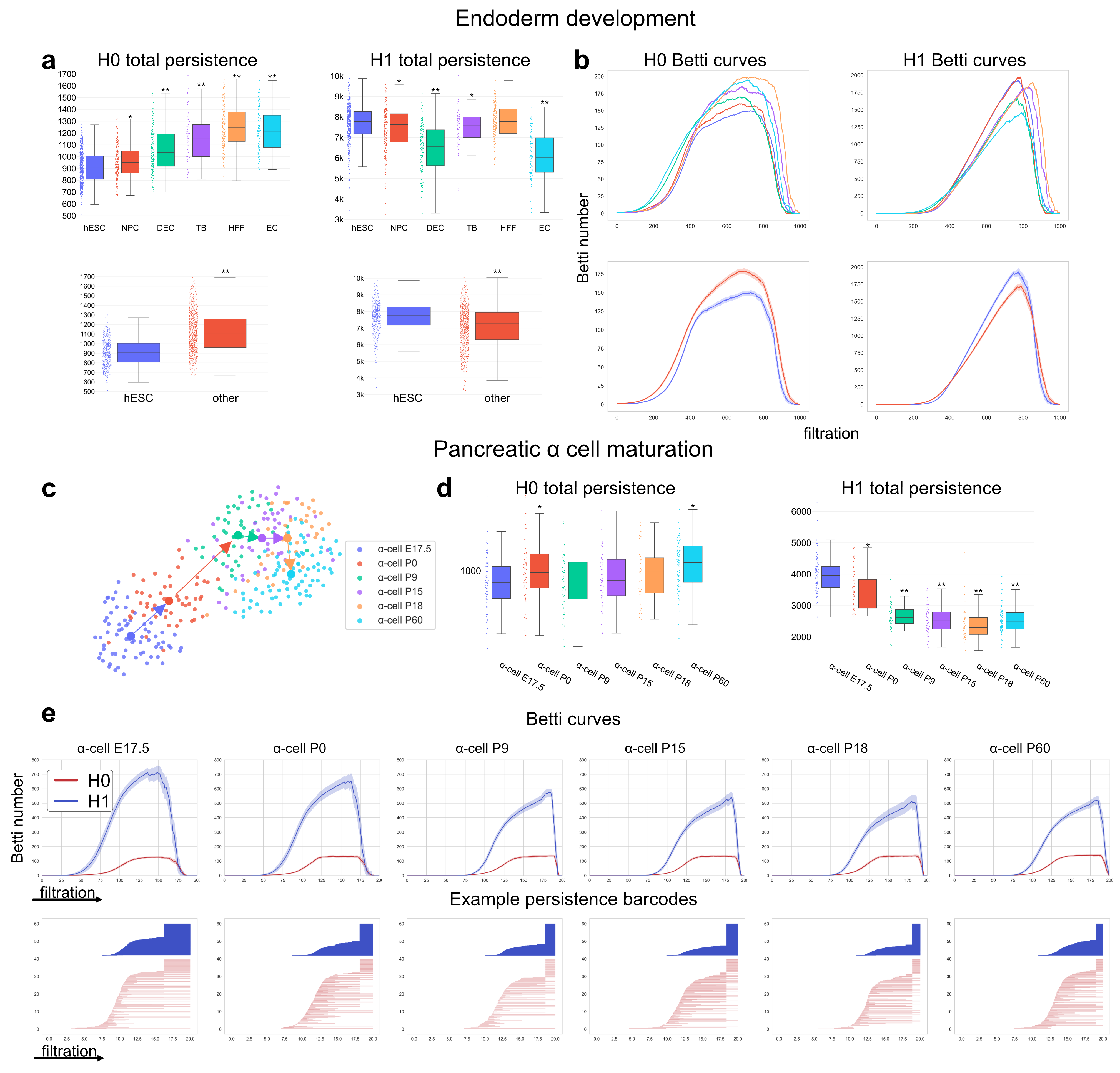}
\centering
\caption{\textbf{Topological analysis of developmental potential.} \textbf{a} For a scRNA-seq data of human definitive endoderm development, the total persistence of $H_0$ and $H_1$ persistence barcodes were computed from a vertex-based clique complex of prior knowledge-based gene network. (hESC: H1 and H9 human embryonic stem cells, NPC: neuronal progenitor cells, DEC: definitive endoderm cells, TB: trophoblast-like cells, HFF: human foreskin fibroblasts, EC: endothelial cells) * and ** indicate $p$-value less than 0.05 and 1e-10, respectively for Wilcoxon tests between hESC and other cell types. \textbf{b} Average $H_0$ and $H_1$ Betti curves for the detailed celltypes and for hESC versus all other cell types. For the latter, 95\% confidence intervals for the mean curve are shown. \textbf{c} A scRNA-seq data of pancreatic $\alpha$ cell maturation where the arrows show the ground truth developmental trajectory. \textbf{d} The total persistence of $H_0$ and $H_1$ persistence barcodes computed from vertex-based clique complex of the prior knowledge-based gene network. * and ** indicate $p$-value less than 0.05 and 1e-10, respectively for Wilcoxon tests between $\alpha$-cell E17.5 and other cell states. \textbf{e} Average Betti curves for each cell state with 95\% confidence intervals of curve mean, and persistence barcodes of example cells from each state.}
\label{fig:potency}
\end{figure}

We first analyzed a scRNA-seq data of human definitive endoderm development\cite{chu2016single}. In this dataset, the human embryonic stem cells (hESC) are pluripotent cells that differentiate into several lineage-specific progenitors. Evaluating the persistent homology of cells at each state, we observe an increase in $H_0$ total persistence and a decrease of $H_1$ total persistence along the differentiating progress (Fig. \ref{fig:potency}a). The $H_0$ and $H_1$ persistent homology captures connected components and loop-like structures which indicates that the gene network of pluripotent cells tends to have less isolated components (shorter $H_0$ persistence) and differentiated cells tend to be active in a localized part of the gene network (shorter $H_1$ persistence). Comparing hESC cells with all other cells also shows significantly lower $H_0$ persistence and higher $H_1$ persistence in hESC cells (Fig. \ref{fig:potency}a). The Betti curves summarize the number of connected components ($H_0$) and $1$-dimensional holes or loops ($H_1$) at every filtration value which also show that hESC cells have fewer disconnected parts and larger-scale loops with more coverage of the gene network(Fig. \ref{fig:potency}b). 

We also analyzed a scRNA-seq data of mouse pancreatic $\alpha$ cell maturation tracking along one cell lineage\cite{qiu2017deciphering}. In this dataset, scRNA-seq experiments were performed for pancreatic $\alpha$ cells at different developmental stages including embryonic data 17.5 (E17.5) and postnatal day P0, P9, P15, P18, and P60 (Fig. \ref{fig:potency}c). During the maturation, we also observed a similar pattern with increasing $H_0$ persistence and decreasing $H_1$ persistence (Fig. \ref{fig:potency}d). The observation is further confirmed in the Betti curves and persistence barcodes of example cells from each maturation stage (Fig. \ref{fig:potency}e). Together, these examples demonstrate that the topological signatures of gene networks reflect the developmental potential of cells.

\subsection{Topoligical machine learning improves cell type classification}

\begin{figure}[h]
\includegraphics[width=0.7\textwidth]{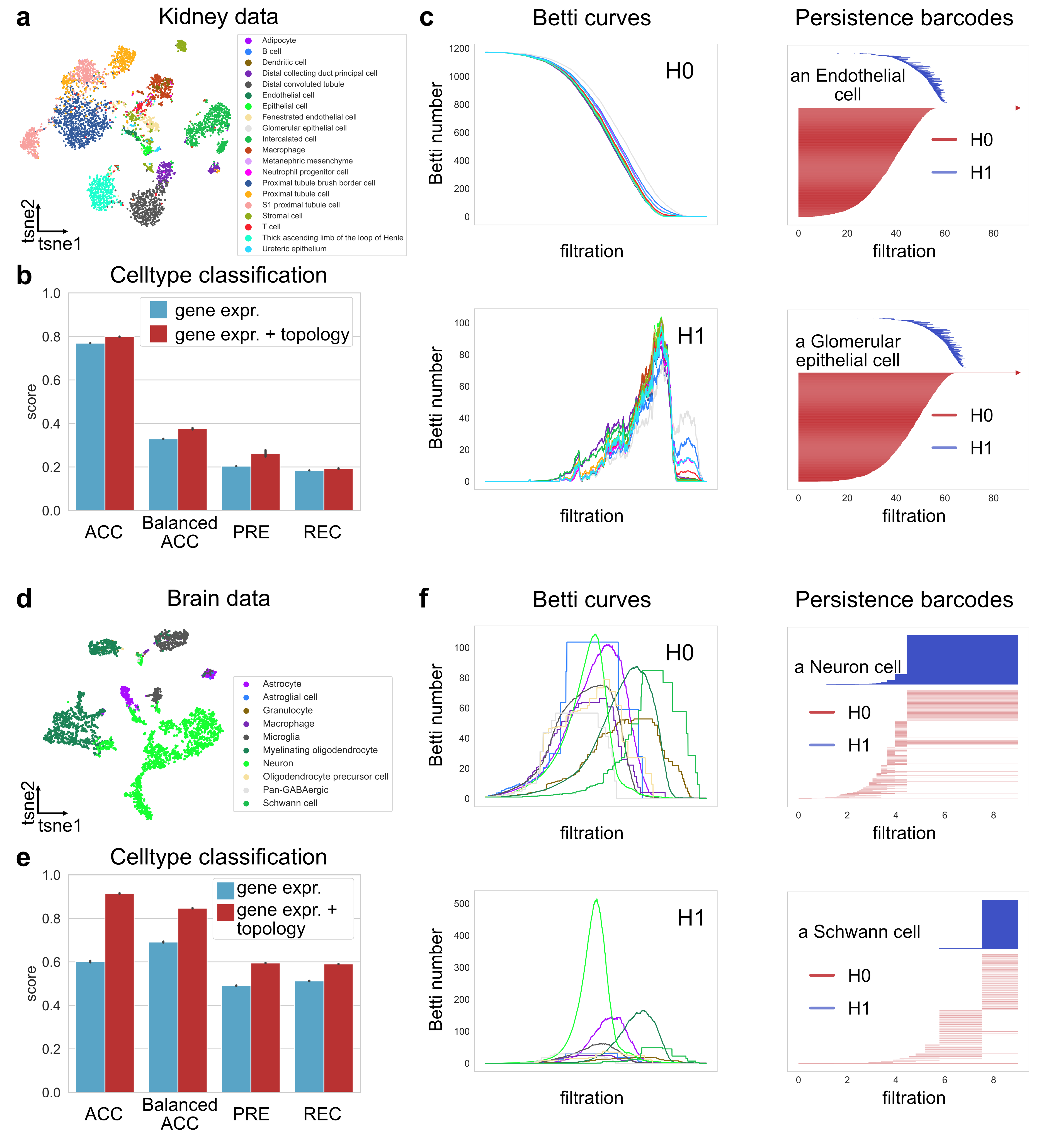}
\centering
\caption{\textbf{Topology-assisted cell type annotation} \textbf{a} A scRNA-seq dataset of Kidney with expert annotated cell types. \textbf{b} The classification performance with or without using topological features. The performance is evaluated by accuracy (ACC), adjusted balanced accuracy (Balanced ACC), precision (PRE) and recall (REC) both with macro-average. \textbf{c} The average $H_0$ and $H_1$ Betti curves for each cell type and persistence barcodes of two example cells for the Vietoris-Rips filtration on the cell-specific gene networks. \textbf{d,e} The classification performances on a scRNA-seq dataset of the brain. \textbf{f} The average $H_0$ and $H_1$ Betti curves for each cell type and persistence barcodes of two example cells for the vertex-based clique complex filtration on the prior knowledge-based gene network.}
\label{fig:celltype}
\end{figure}

Having shown the utility of topological and geometric structures in single-cell data for analyzing transition cells and developmental potential, here, we explore the usage of these structures in the general task of cell type annotations. We used a mouse brain dataset and a mouse kidney dataset from the cell type annotation subtask in a benchmarking resource\cite{ding2022dance} with pre-defined train/test splittings. In this task, a predictive model is trained on annotated data to predict cell types from their gene expression profiles. 

We performed two topological characterizations of the gene networks for each cell. First, a cell-specific gene network (CSN) is constructed using a correlation-based approach\cite{dai2019cell} which results in an edge-weighted gene network for each cell. The CSNs were generated on processed data using the preprocessing pipeline of SingleCellNet\cite{tan2019singlecellnet} to select the marker genes of each cell type. Then, we computed persistent homology using an inverse Vietoris-Rips complex-based filtration which adds $1$-simplexes and subsequently the higher dimensional simplices with large edge weights first. Second, we utilize a prior knowledge-based gene network (SCENT\cite{banerji2013cellular}) on the single-cell datasets without gene filtering. The same gene network structure is assigned to each cell but with different node weights assigned from gene expression levels. In this case, persistent homology was computed based on a vertex-based clique-complex filtration where $0$-simplices and subsequent higher dimensional simplices with higher weights are added first. The total persistence and persistence entropy\cite{atienza2020stability} of the resulting persistence diagrams were used as topological features for the cells.

In both benchmarks, the classification performance is improved with the additional topological features evaluated by accuracy, balanced accuracy, and macro-average precision and recall (Fig. \ref{fig:celltype}. The Betti curves and persistence barcodes of several example cells demonstrate the differences in the topological signature of gene networks across different cell types (Fig. \ref{fig:celltype}c,f). In the brain dataset, interestingly, we observe significantly longer persistence in the $H_1$ persistence barcodes of neuron cells compared to other brain cells indicating a broader coverage of gene network and diverse functions. This observation agrees with the diverse signals sent and functions controlled by neuron cells\cite{schiapparelli2022proteomic}.

\section{Methods}\label{sec:method}
\subsection{Curvature on graphs}\label{sec:graphcurvature}
Given a metric, the Ricci curvature describes how much the local geometry on a manifold differs from that of the ordinary Euclidean space. It measures intrinsic local properties of manifolds such as divergence of geodesics and meeting probabilities of coupled random walks. Several constructions have been introduced to define Ricci curvature on graphs, such as Ollivier-Ricci\cite{lott2009ricci,ollivier2009ricci,ollivier2010survey} curvature. For an edge connecting two nodes, Ollivier-Ricci curvature (ORC) measures the difference between the edge distance between the nodes and the optimal transport distance between the nodes' neighborhoods. Let $G=(V,E)$ be an undirected graph with vertices $V=\{v_i\}_{i=1}^n$ and edges $E=\{e_{ij}\}_{1\leq i,j\leq n}$, a measure is defined for each node describing its local neighborhood such that 
\begin{equation*}
m_{v_i}(v_j; \alpha)=\begin{cases}
    \alpha,\,\mathrm{if}\, j=i, \\
    (1-\alpha)/|N(v_i)|, \,\mathrm{if}\, i\neq j\in N(v_i), \\
    0, \,\mathrm{elsewhere},
\end{cases}
\end{equation*}
where $N(v_i)$ is the set of nodes connected to $v_i$ in $G$ and $\alpha\in [0,1]$ is the parameter annotating the weight on the center node. The ORC $\kappa_{ij}$ between $v_i$ and $v_j$ is then defined to be
\begin{equation*}
    \kappa^{\mathrm{orc}}_{ij} = 1-\frac{d_W(m_{v_i},m_{v_j})}{d(v_i,v_j)},
\end{equation*}
where $d()$ is the shortest path distance on $G$ and $d_W$ is the Wasserstein distance with $d()$ as the ground metric. Specifically, an optimal transport problem is solved, $d_W(m_{v_i},m_{v_j})=\inf\limits_{\pi\in\Pi(m_{v_i},m_{v_j})}<\pi,C>_F$ with $C_{ij}=d(v_i,v_j)$ and $\Pi(a,b)=\{\pi\in\mathbb{R}_+^{n\times n}|\pi\mathbf{1}=a,\pi^T\mathbf{1}=b\}$.

\subsection{Persistent homology}\label{sec:persistenthomology}
Persistent homology\cite{edelsbrunner2002topological,zomorodian2004computing,edelsbrunner2008persistent} provides a comprehensive multiscale topological characterization by tracking the appearance and disappearance of homology groups through a filtration which is a growing sequence of simplicial complexes defined on the data. An abstract $k$-simplex is a set of $k+1$ vertices denoted $\sigma^k=\{v_0,\cdots,v_{k}\}$. Any subset $\sigma'\subseteq\sigma^k$ is called a face of $\sigma^k$. A simplicial complex $K$ is a set of simplices satisfying that all faces of any simplex in $K$ is also in $K$ and the intersection of any pair of simplices is either empty or a common face of the two. A filtration of a simplicial complex $K$ is a nested sequence of its subcomplexes, $\emptyset=K^0\subset K^1\subset\cdots\subset K^n=K$. A $k$-chain of a simplicial complex $K$ denoted by $c_k$ is a formal sum of $k$-simplices in $K$ with coefficients from a chosen set, for example, $\mathbb{Z}_2$. The $k$-chains of $K$ forms a group called the $k$th chains group denoted by $C_k(K)$. The $k$-chains are connected by a linear boundary operator $\partial_k: C_{k}(K)\rightarrow C_{k-1}(K)$ such that $\partial_k(\sigma^k)=\sum\limits_{i=0}^{k}(-1)^i\hat{\sigma}^{k-1}_i$ where $\hat{\sigma}^{k-1}_i$ is a face of $\sigma^k$ obtained by removing vertex $i$. Based on the boundary operator, two groups are defined, the kernel of $\partial_k$, $Z_k(K)=\mathrm{ker}(\partial_k)$ whose elements are called $k$-cycles and the image of $\partial_{k+1}$ denoted by $B_k(K)$ also called the $k$th boundary group. The $k$th homology group is then defined as the quotient group $H_k(K)=Z_k(K)/B_k(K)$ whose rank $\mathrm{rank}(H_k(K))=\mathrm{rank}(B_k(K))-\mathrm{rank}(B_k(K))$ is also the $k$th Betti number of $K$ representing the number of $k$-dimensional holes in $K$. On the filtration of $K$, the $p$-persistent $k$th homology group is defined to be $H_k^p(K^i)=Z_k(K^i)/(B_k(K^{i+p})\cap Z_k(K^i))$ which intuitively represents a topological feature observed at filtration step $i$ and persists through step $i+p$. An equivalence class in $H_k(K^i)$ persisting through $H_k(K^j)$ but is not present in $H_k(K^{i-1})$ or $H_k(K^{j+1})$ result in a persistence pair conveniently represented as the interval $[x_i,x_j)$ often called the birth-death pairs where $x_i$ and $x_j$ are the filtration values of $K^i$ and $K^j$ respectively. Persistent homology characterization of a dataset results in a collection of such birth-death pairs and is often visualized as persistence barcodes (plotting each pair as a horizontal bar whose two endpoints correspond to the birth and death values) or persistence diagrams (plotting each pair as a point in 2D).

\subsection{Curvature in single-cell data}
The raw single-cell data was first preprocessed by normalizing total counts in every cell and log1p transformation ($\log(1+x)$). Principal component analysis is then performed with the selected highly variable genes. A $k$-nearest neighbor graph is constructed based on the Euclidean distance in the space of top principal components. The ORC is computed on this cell network with the $\alpha$ parameter (the portion of mass assigned to the center cell when determining a mass distribution representing the neighborhood of the cell) set to 0.5. The preprocessing was performed using the Scanpy package\cite{wolf2018scanpy}.

\subsection{Topology of cell networks}
To characterize the topological structures of the cell network for each cell, we use two approaches, a local persistent homology and a relative persistent homology. The local persistent homology aims to capture the local structure surrounding a cell in the cell network. Here, we adopt a simple approach by computing the regular persistent homology on a sub-network surrounding a cell defined by either the top $k$ nearest neighbors or a distance cutoff. This approach has been shown effective in capturing local topological structures in various applications such as biomolecular structure analysis\cite{meng2020weighted}. The relative persistent homology, on the other hand, captures the significance of a cell or the neighborhood of a cell in assembling the global structure of the dataset. Let $\emptyset=K^0\subset K^1\subset\cdots\subset K^n=K$ be a filtration of the whole dataset. For cell $i$, we define a subcomplex $L_i$ which contains only this cell or its local neighborhood on the network. Then, relative persistent homology examines the homology on the relative chain groups which are quotient groups $C_k(K^j)/C_k(K^j\cap L_i)$. The impact of cell $i$ on assembling the global dataset structure is then quantified by computing the Wasserstein distance between the relative persistent homology diagrams and the regular persistent homology diagram of the whole dataset. For both approaches, we used the Vietoris-Rips filtration with the Euclidean distance between cells in their PCA embeddings.

\subsection{Topology of gene networks}
Here, we consider two types of gene networks, a cell-specific gene network in the form of edge-weighted networks and a prior knowledge-based gene network in the form of node-weighted networks. 

The package CSN\cite{dai2019cell} is used to construct a cell-specific gene network for each cell on the top highly variable genes in the dataset. The core statistic in this method evaluates the local association between each gene pair for every cell. Specifically, for cell $k$ and genes $x$ and $y$, $\rho_{xy}^{(k)}=n_{xy}^{(k)}/n-(n_x^{(k)}/n)(n_y^{(k)}/n)$, where $n$ is the total number of cells. The parameter $n_x^{(k)}$ is predefined and induces an interval $I_x^{(k)}$ of the expression of gene $x$ in the top $n_x^{(k)}$ cells whose gene $x$ expression is the closest to cell $k$. Similarly, an interval for gene $y$, $I_y^{(k)}$ is determined by $n_y^{(k)}$. Then $n_xy^{(k)}$ counts the number of cells in whose expressions of gene $x$ and gene $y$ both fall in the two intervals, respectively. Here, we used the top 1000 highly variable genes and used the default parameter values in the CSN method\cite{dai2019cell} with $n_x^{(k)}=n_y^{(k)}=0.1n$ and significant level set to $0.01$. The constructed cell-specific gene networks are edge-weighted networks. Denoting the network of a cell by $G=(V,E,W^{(e)})$, we compute persistent homology with the Vietoris-Rips filtration 
$$VR(\delta)=\{\sigma : \forall \sigma^{(1)}=(i,j)\subseteq\sigma, W^{(e)}_{ij}\geq\delta \,\,\mathrm{and}\,\, (i,j)\in E\}.$$ 
The filtration is computed from $\delta=\delta_{max}=\max\{W^{(e)}\}$ to $\delta=0.$
The resulting persistence pairs $[b_i,d_i)_i$ are transformed to $[\delta_{max}-b_i,\delta_{max}-d_i)_i$.

For the prior knowledge-based gene network, a base network is first assigned to each cell and the gene expression levels in each cell are assigned as node weights. Denoting the network of a cell by $G=\{V,E,W^{(v)}\}$, persistent homology is computed using the vertex-based clique complex filtration 
$$CL(\delta)=\{\sigma : \forall \sigma^{(0)}_i\subseteq \sigma, W^{(v)}_i\geq\delta ; \forall \sigma^{(1)}=(i,j)\subseteq\sigma, \min\{W^{(v)}_i, W^{(v)}_j\}\geq\delta \,\,\mathrm{and}\,\,(i,j)\in E\}.$$ 
Similar to the edge-weighted network, the filtration is computed from $\delta=\delta_{max}=\max\{W^{(v)}\}$ to $\delta=0$. The resulting persistence pairs $[b_i,d_i)_i$ are also transformed to $[\delta_{max}-b_i,\delta_{max}-d_i)_i$. The log1p transformed gene expression levels are used to assign node weight and the full dataset without gene filtering is used with the knowledge-based gene network.

The computation of filtration and persistent homology was based on the packages Gudhi\cite{gudhi:urm}, Dionysus2\cite{dionysus2}, and Ripser\cite{Bauer2021Ripser}.

\subsection{Featurization of persistence diagrams}

For dimension $k$, persistent homology computation results in a collection of $n^{(k)}$ persistence pairs $[b^{(k)}_i,d^{(k)}_i)_{i=1}^{n^{(k)}}$ where $b^{(k)}_i$ and $d^{(k)}_i$ are the filtration values corresponding to the birth and death of a topological feature ($k$-dimensional holes). Several summaries and features are derived from the persistence pairs. Total persistence describes the overall significance of topological features in the data and is computed as $L^{(k)}=\sum_{i=1}^n (\bar{d}^{(k)}_i-b^{(k)}_i)$ where $\bar{d}^{(k)}_i=\min\{d^{(k)}_i, \delta_{max}\}$. Persistence entropy\cite{atienza2020stability} describes the heterogeneity of persistence similar to Shannon entropy and is stable with respect to small perturbations in the input space. Specifically, it is computed as $E^{(k)}=-\sum_{i=1}^{n^{(k)}}\frac{l^{(k)}_i}{L^{(k)}}\log_2(\frac{l^{(k)}_i}{L^{(k)}})$ where $l^{(k)}_i=\bar{d}^{(k)}_i-b^{(k)}_i$ is the persistence of the $i$th pair. In addition to global summaries, Betti curves describe the structural changes along the filtration and are convenient for illustrating the average behavior of a group of persistence barcodes. For a filtration value $\delta$, the $k$-dimensional Betti curve is computed as $BC^{(k)}(\delta)=\#\{[b^{(k)}_i,d^{(k)}_i):\delta\in[b^{(k)}_i,d^{(k)}_i)\}$. Betti curve is computed on a discretization of the filtration interval $[0,\delta_{max}]$. The Gudhi package\cite{gudhi:urm} was used to compute the persistence entropy and Betti curves.

\subsection{Machine learning and evaluation metrics}

In the application of cell type classification, the implementation of random forest model in scikit-learn package\cite{scikit-learn} was used with 5000 trees and "class\_weight" parameter set to "balanced". All other parameters are set to default values. The classification results on the testing set were evaluated using accuracy: $(1/N)\sum_i1(\hat{y}_i=y_i)$; balanced accuracy (adjusted): $\left(1/\sum_i\frac{1}{N_{c(i)}}\right)\sum_i1(\hat{y}_i=y_i)\frac{1}{N_{c(i)}}-\frac{1}{C}$; precision (macro-average): $\frac{1}{C}\sum_{c=1}^C\frac{|\hat{Y}_c\cap Y_c|}{|\hat{Y}_c|}$; and recall (macro-average): $\frac{1}{C}\sum_{c=1}^C\frac{|\hat{Y}_c\cap Y_c|}{|Y_c|}$. Here, $N$ is the total number of samples, $N_{c(i)}$ is the number of samples of the same class as sample $i$ in the ground truth, $y_i$ is the true label of sample $i$, $\hat{y}_i$ is the predicted label of sample $i$, $C$ is the number of classes, $Y_c$ is the set of samples of class $c$ in ground truth, $\hat{Y}_c$ is the set of sample predicted to be class $c$, and $1()$ is the indicator function.

\section{Conclusion}
To exploit the underlying complex structures in scRNA-seq data, we developed scGeom, a tool to derive topological and geometric signatures from the network of cells and gene networks associated with each cell. It utilizes Ollivier-Ricci curvature, local persistent homology, relative persistent homology and persistent homology filtrations for edge-weighted and node-weighted networks. The utilities of these structural characterizations have been demonstrated on real scRNA-seq datasets for identifying transition cells, quantifying pluripotency or developmental potency of cells, and assisting in the classification of cells.

Persistent homology is used as a structural descriptor in this work without tracing back to individual cells or genes from the topological signatures. Recently, several methods were proposed to connect the topological features with the input data\cite{bendich2020stabilizing,obayashi2018volume,cang2020persistent} which could help interpret the topological features in terms of cells or genes. The topological structures are compared as topological summaries between large-scale gene networks in this work. When comparing small-scale networks in the future, such as specific pathways, two networks could have the same structure but different arrangements of genes which will result in identical topological structures. This could be addressed by a very recent work that compares topological summaries while considering the differences in the original data\cite{yoon2023persistent}. 

This work presents one of the initial endeavors to apply persistent homology to scRNA-seq data. The methods are also potentially applicable to other single-cell omics data\cite{baysoy2023technological} given some similarity measurement between cells or association scores between features. With the recent developments of multiparameter persistent homology\cite{botnan2022introduction}, different metrics can be considered simultaneously in complex data such as single-cell multi-omics data with multiple similarity measurements and spatial transcriptomics data with both spatial distance and gene expression similarities.

\section{Data and code availability}
All datasets used are publicly available. 1) The myelopoiesis data\cite{olsson2016single} is available on GEO with accession number GSE70245; 2) The iPSC data\cite{bargaje2017cell} is available in the Supplementary Data of the original publication; 3) The endoderm development data\cite{chu2016single} is available on GEO with accession number GSE75748; 4) The pancreatic $\alpha$ cell data\cite{qiu2017deciphering} is available on GEO with accession number GSE87375; 5) The mouse brain and kidney data with annotated cell types and predetermined train/test splits were downloaded using the Dance package\cite{ding2022dance}. 

The package scGeom is available at https://github.com/zcang/scGeom.

\section{Acknowledgements}
TH is partly supported by Center for Research in Scientific Computing at NC State University. ZC is partly supported by a start-up grant from the NC State University and an NSF grant DMS-2151934.


\end{document}